\documentclass[a4paper,reqno,twoside,final,psamsfonts]{amsart}
\usepackage[latin1]{inputenc}
\usepackage[T1]{fontenc}
\usepackage{textcomp}
\usepackage[british]{babel}
\usepackage{newcent}
\usepackage[pagebackref]{hyperref}
  \hypersetup{%
   colorlinks=true, anchorcolor=black, citecolor=magenta, urlcolor=blue, linkcolor=red,
    pdftitle= {Zeno Dynamics of von Neumann Algebras},
    pdfauthor= {Andreas U. Schmidt},
    pdfsubject= {2000 Mathematical subject classification: 46L60; 81P15, 81R15; PACS: 03.65.Xp, 03.65.Db, 02.30.Tb},
    pdfkeywords= {Quantum Zeno effect, von Neumann algebra, modular group} }
\newif\ifpdf\pdftrue
\providecommand{\href}[2]{#2}
\usepackage{xspace}
\usepackage{mathrsfs} 
\usepackage{amsmath}
\usepackage{amssymb}
\usepackage{amsthm}
\usepackage{amsfonts}
\newcommand{\emox}[1]{%
{
\ensuremath{\mathord{#1}}
}\xspace
}
\newcommand{\CC}{{\ensuremath{\mathord{\mathbb{C}}}}\xspace}
\newcommand{\RR}{{\ensuremath{\mathord{\mathbb{R}}}}\xspace}

\newcommand{\NN}{{\ensuremath{\mathord{\mathbb{N}}}}\xspace}

\DeclareMathAlphabet{\Bmi}{OT1}{cmm}{b}{it}

\newcommand{\HS}{{\ensuremath{\mathord{\mathcal{H}}}}\xspace} 

\newcommand{\EPS}{\emox{\varepsilon}} 
\newcommand{\IM}{\operatorname{Im}}

\newcommand{\wlim}{\mathop{\operatorname{w-lim}}}
\newcommand{\slim}{\mathop{\operatorname{s-lim}}}

\newcommand{\ABS}[1]{{\ensuremath{\mathord{\left|#1\right|}}}}
\newcommand{\NORM}[1]{{\mathord{\left\|#1\right\|}}}
\newcommand{\DEF}{
  \mathbin{\smash[t]{\overset{\scriptscriptstyle\mathrm{def}}{=}}}}

\newcommand{\CDOT}[1][]{\mathbin{\widehat{\cdot}_{#1}}}

\newcommand{\SPROD}[2]{
            \ensuremath{\mathord{\left\langle#1\,\mathord{,}\,#2\right\rangle}}\xspace}

\newcommand{\dd}{\ensuremath{\mathrm{d}}} 
\newcommand{\ee}{\ensuremath{\mathrm{e}}} 
\newcommand{\ie}{, {\it i.e.},\xspace}
\newcommand{\eg}{, {\it e.g.},\xspace}
\newcommand{\GLQQ}{\char"12\kern .08em} 
\newcommand{\GRQQ}{\kern .08em\char"10\xspace}
\def\SetConstructor#1#2#3#4{%
  \def\test{#4}\ifx\test\empty{\ensuremath{\mathord{#1}_{#2}^{#3}}\xspace}%
  \else{\ensuremath{\mathord{#1}_{#2}^{#3}(#4)}\xspace}\fi}
\theoremstyle{plain}
\newtheorem{theo}{Theorem}[section]
\newtheorem{defitheo}[theo]{Definition and Theorem}
\newtheorem*{defitheo*}{Definition and Theorem}
\newtheorem{prop}[theo]{Proposition}
\newtheorem*{prop*}{Proposition}
\newtheorem{coro}[theo]{Corollary}
\newtheorem*{coro*}{Corollary}
\newtheorem{lemm}[theo]{Lemma}
\newtheorem*{lemm*}{Lemma}
\theoremstyle{definition}
\newtheorem{defi}[theo]{Definition}
\newtheorem*{defi*}{Definition}
\newtheorem{defirem}[theo]{Definition and Remark}
\newtheorem*{defirem*}{Definition and Remark}
\newtheorem{rem}[theo]{Remark}
\newtheorem*{rem*}{Remark}
\theoremstyle{remark}

\newcommand{\labelT}[1]{\label{theo:#1}}
\newcommand{\labelP}[1]{\label{prop:#1}}
\newcommand{\labelL}[1]{\label{lemm:#1}}
\newcommand{\labelD}[1]{\label{defi:#1}}
\newcommand{\labelC}[1]{\label{coro:#1}}
\newcommand{\labelR}[1]{\label{rem:#1}}
\newcommand{\labelDR}[1]{\label{defirem:#1}}

\newcommand{\labelE}[1]{\label{eq:#1}}
\newcommand{\refC}[1]{Corollary~\ref{coro:#1}}

\newcommand{\refT}[1]{Theorem~\ref{theo:#1}}

\newcommand{\refL}[1]{Lemma~\ref{lemm:#1}}

\newcommand{\refE}[1]{Equation~\eqref{eq:#1}}
\newenvironment{Ltheo}[2][]{\begin{theo}[#1]\labelT{#2}}{\end{theo}}

\newenvironment{Llemm}[2][]{\begin{lemm}[#1]\labelL{#2}}{\end{lemm}}

\newenvironment{Lcoro}[2][]{\begin{coro}[#1]\labelC{#2}}{\end{coro}}

\newcommand{\OA}{\emox{\mathcal{A}}}
\newcommand{\SC}{\emox{\overline{S}}}
\begin{document}
%
%
\title{Zeno Dynamics of von Neumann Algebras}
\author[A.\ U.\ Schmidt]{Andreas U.\ Schmidt}
\date{17th~August~2002}
\curraddr{Dipartimento di Fisica E.\ Fermi\\
  Universit\`{a} di Pisa\\
  via Buonarroti~2\\
  56127 Pisa PI\\
  Italy}
\address{Fachbereich Mathematik\\
  Johann Wolfgang Goethe-Universität\\
  60054 Frankfurt am Main, Germany\ifpdf\\
  \href{http://www.math.uni-frankfurt.de/~aschmidt}{Homepage}\else\fi}
  \email{\href{mailto:aschmidt@math.uni-frankfurt.de}{aschmidt@math.uni-frankfurt.de}}
\subjclass{\href{http://www.ams.org/msc/46Lxx.html}{46L60}; 
\href{http://www.ams.org/msc/81Pxx.html}{81P15}, 
\href{http://www.ams.org/msc/81Rxx.html}{81R15}}
\thanks{\emph{PACS Subject Classification.} 
\href{http://www.aip.org/pacs/pacs01/pacs0100-02.htm}{03.65.Xp, 03.65.Db, 02.30.Tb}}
\thanks{This research was supported by a research grant from the
Deutsche Forschungsgemeinschaft 
\href{http://www.dfg.de}{DFG}. The author wishes to thank the 
\href{http://www.df.unipi.it}{Dipartimento di Fisica E.~Fermi}, 
\href{http://www.unipi.it}{Universit\`{a} di Pisa}, and the
\href{http://www.infn.it}{INFN} for their hospitality.
Heartfelt thanks go to  
Giovanni Morchio 
(Pisa, Italy), 
Saverio Pascazio and Paolo Facchi (Bari, Italy),
Daniel Lenz (Chemnitz, Germany),
and Matthias Schork (Frankfurt am Main, Germany) 
for many helpful hints and discussions. The suggestions
of the referees are gratefully acknowledged.}
\begin{abstract}
The dynamical quantum Zeno effect is studied in the
context of von Neumann algebras. We identify a localized 
subalgebra on which the Zeno dynamics acts by automorphisms.
The Zeno dynamics coincides with the modular dynamics
of that subalgebra, if an additional assumption is satisfied. 
This relates the modular operator of that subalgebra to the 
modular operator of the original algebra by a variant of the 
Kato--Lie--Trotter product formula.
\end{abstract}
\maketitle
%
%
\section{Introduction}
The Zeno or ``a watched pot never boils'' effect has, over the last decades,
attracted a lot of interest from quantum physicists, 
see~\cite{FGMPS00,FNP01,GUS02,HW86,HW98,HNNPR98,WBT01} and the extensive 
list of references in~\cite{NNP96}. The effect, which consists in
the possible impedition of quantum evolution\eg decay processes,
 under the influence of frequent measurement events, or, more generally, 
frequent system-environment interactions, has even entered popular science 
texts~\cite{b:WIC95}. It is a striking example of the peculiarities of 
quantum theory whose origin can be traced back to the geometry of the Hilbert 
space~\cite{AA90}, which implies a quadratic short-time behavior of
 transition probabilities~\cite{PL98}. 

Here, we rely on the mathematical formulation of the strict Zeno paradox as
presented by Misra and Sudarshan in~\cite{MS77}\ie the limit of infinitely 
frequent measurements, which we will now briefly sketch. Given the Hamiltonian
evolution $U(t)=\ee^{-iHt}$ on the Hilbert space \HS, 
with $H$ lower semibounded, and a projection $E$ on \HS, one assumes
the existence of the limits
\[
W(t)\DEF\slim_{n\rightarrow\infty}\bigl[EU(t/n)E\bigr]^n
\]
in the strong sense on \HS, for all $t\in\RR$ (this condition can
be relaxed to $t\geq0$ if a CPT-symmetry is present). It is then shown,
using methods of complex analysis,
that $W(t)$ is a strongly continuous group, $W(t+s)=W(t)W(s)$, for all
$t\in\RR$, and $W(-t)=W(t)^\ast$. In particular one finds a lower semibounded
operator $B$ and a projection $G$ such that $BG=GB=B$ which induces
the Zeno dynamics: $W(t)=G\ee^{-iBt}G$. If one employs the initial condition
$\slim_{t\rightarrow0_+}W(t)=E$, one can identify $G$ with $E$. Thus one
sees that the Zeno dynamics is a modified Hamiltonian dynamics, which
is confined to the Zeno subspace $E\HS$.

Although the limit of infinitely frequent measurements has been argued to 
be unphysical due to Heisenberg uncertainty~\cite{NNP96}, 
it is still of conceptual 
interest. This is in particular the case if one wants to study the induced
limiting dynamics $W(t)$ on the Zeno subspace and to compare it
with the original one on the full space. This has, however, only been done for
simple examples~\cite{FGMPS00,FPSS01}. It turns out that in these 
examples, the Zeno dynamics corresponds to Hamiltonian evolution
with additional constraints and boundary conditions. For example, one finds
an infinite well potential term, corresponding to Dirichlet boundary conditions, 
when the projector is multiplication with
the characteristic function of an interval in the one-dimensional case.

In this paper we show how the treatment of~\cite{MS77} can be carried over
to the modular flow of a von Neumann algebra $\OA$. First and foremost, this
is a direct generalization of the result of Misra and Sudarshan to dynamics
whose generators are not lower semibounded.
The special role projections play for von Neumann algebras
gives our generalization some additional impact, in view of the ongoing
discussion over the projection postulate. Furthermore, our analysis can also 
straightforwardly be applied to KMS-states of $W^\ast$-dynamical systems
for inverse temperatures $0<\beta\leq\infty$.
But what is, in our view, most important, is that our treatment yields
an explicit identification of the Zeno dynamics: It can, in favorable cases,
be shown to coinced with the unique modular
flow of the localized von Neumann subalgebra $E\OA E$. This result can be 
viewed as a variant of the Kato--Lie--Trotter product 
formula~\cite[Corollary~3.1.31]{b:BR79/81}.

In the following section, we present this generalization of the
strict Zeno paradox. The final section contains some remarks on
the status of the main \refT{main}, its weaknesses and possible
extensions, as well as an outlook towards physical applications.
\section{The Zeno Paradox in the Context of von Neumann Algebras}
\begin{Ltheo}{main}
Let \OA be a von Neumann algebra with faithful, normal state
$\omega$, represented on the Hilbert space 
$\HS$ with cyclic and separating vector $\Omega$ associated with $\omega$.
Let $\Delta$ be the modular operator of $(\OA,\Omega)$. Let $E\in\OA$ be 
a projection. 
Set $\OA_E\DEF E\OA E$, and define a subspace of \HS by 
$\HS_E\DEF\overline{\OA_E\Omega}\subset E\HS$.
Assume:
  \begin{enumerate}
  \item\label{ass1}
For all $t\in\RR$, the strong operator limits 
\[
W(t)\DEF\slim_{n\rightarrow\infty}
\bigl[E \Delta^{i t/n} E\bigr]^n
\]
exist, are weakly continuous in $t$, and satisfy the 
initial condition 
\[
\wlim_{t\rightarrow0}W(t)=E.
\]
  \item For all $t\in\RR$, the following limits exist:
\[
W(t-i/2)\DEF\slim_{n\rightarrow\infty}
\bigl[E \Delta^{i (t-i/2)/n} E\bigr]^n,
\]
where the convergence is strong on the common, dense
domain $\OA\Omega$.
\end{enumerate}
Then the $W(t)$ form a strongly continuous group of unitary operators
on  $\HS_E$. The group $W(t)$
induces an automorphism group $\tau^E$ of  
$\OA_E$ by
\[
\tau^E\colon\OA_E\ni A_E\longmapsto \tau^E_t(A_E)\DEF W(t)A_EW(-t)=
W(t)A_EW(t)^\ast,
\]
such that $(\OA_E,\tau^E)$ is a 
$W^\ast$-dynamical system. The vectors 
$W(z)A_E\Omega$,  $A_E\in\OA_E$, are holomorphic in the strip $0<-\IM{z}<1/2$ and 
continuous on its boundary.
\end{Ltheo}Notice that $\OA_E$ is a von Neumann subalgebra of 
$\OA$, see~\cite[Corollary~5.5.7]{b:KR83/86}, for which  $\Omega$ is
cyclic for $\HS_E$ and separating. Thus, $\Omega$ induces a faithful
representation of $\OA_E$ on the closed
Hilbert subspace $\HS_E$, and thus all notions above are well-defined.
\begin{Lcoro}{modular-identify}
  Let $\Omega_E$ be a vector in 
$\HS_E$ which induces a faithful, normal state $\omega_E$ on $\OA_E$, and
denote by $\Delta_E$ the modular operator of $(\OA_E,\omega_E)$.
Assume further the validity of the following additional condition:
\begin{enumerate}
\setcounter{enumi}{2}
\item  
For all $A_E$, $B_E\in\OA_E$ holds
\[
\lim_{t\rightarrow0}\SPROD{W(-t-i/2)A_E\Omega_E}{W(t-i/2)B_E\Omega_E}=
\SPROD{\smash{\Delta_E^{1/2}}A_E\Omega_E}{\smash{\Delta_E^{1/2}} B_E\Omega_E}.
\]
\end{enumerate}
Then $\tau^E$ is the modular atuomorphism group of $(\OA_E,\Omega_E)$.
\end{Lcoro}
The remainder of this section contains the proof
of the above theorem and its corollary, split into several lemmata.
In all these, we will only use conditions~i) and~ii).
Only after that will condition~iii) be used to identify
the modular group.
 
Set $S\DEF\bigl\{z\in\CC\bigm|-1/2<\IM z<0\bigr\}$.
Define operator-valued functions
\[
F_n(z)\DEF \bigl[E \Delta^{i z/n} E\bigr]^n, \quad
\text{for }z\in\SC,\ n\in\NN.
\]
The $F_n(z)$ are operators whose domains of definition contain
the common, dense domain $\OA\Omega$. They depend holomorphically on $z$ 
in the sense that the vector-valued functions $F_n(z)A\Omega$ are 
holomorphic on $S$ and continuous on \SC for every $A\in\OA$. For this
and the following lemma see~\cite[Section~2.5, Section~5.3, 
and Theorem~5.4.4]{b:BR79/81}.
\begin{Llemm}{FnBound}
  For $z\in\SC$ and $\Psi\in D(\Delta^{\ABS{\IM z}})$ holds the estimate
\[
\NORM{F_n(z)\Psi}\leq\NORM{\Psi},
\]
for all $n\in\NN$.
\end{Llemm}
\begin{proof} 
Define vector-valued functions 
$
f_k^{\Psi,n}(z)\DEF \bigl[E\Delta^{iz/n}E\bigr]^k \Psi.
$
These are well-defined for $z\in\SC$, $\Psi\in D(\Delta^{\ABS{\IM z}})$ and
all $k\leq n$, since for such $\Psi$, $z$ we have
$\bigl[E\Delta^{iz/n}E\bigr]^{k-1}\in D(E\Delta^{iz/n}E)$.
Approximate $f_{k-1}^{\Psi,n}(z)$ by elements of the form $A_l\Omega$, $A_l\in\OA$.
Then for any $B\in\OA$ holds
\begin{align*}
  \ABS{\SPROD{\smash{B\Omega}}{\smash{E\Delta^{iz/n}E A_l  \Omega}}} &=
  \ABS{\SPROD{\Omega}{\smash{B^\ast E \Delta^{iz/n}E A_l 
    \Delta^{-iz/n}\Omega}} } \\ &
 =\ABS{\omega(B^\ast E\sigma_{z/n}(E A_l)) }\\ 
&\leq \NORM{B^\ast E\Omega}\NORM{EA_l\Omega} \\ &\leq
  \NORM{B}\NORM{A_l\Omega}.
\end{align*}
Here, $\omega$ is the state on \OA associated with the cyclic and separating
vector $\Omega$ (we always identify elements of \OA with their representations
on \HS), and $\sigma$ denotes the modular group. The first estimate above follows explicitly
from the corresponding property of $\sigma$, see~\cite[Proposition~5.3.7]{b:BR79/81}
(the connection between faithful states of von Neumann algebras and
KMS states given by Takesaki's Theorem~\cite[Theorem~5.3.10]{b:BR79/81} is 
used here and in the following).
This means $\NORM{E\Delta^{iz/n}E A_l\Omega}\leq\NORM{A_l\Omega}$, and since 
$A_l\Omega\longrightarrow f_{k-1}^{\Psi,n}(z)$ in the norm of \HS, it follows
$\NORM{\smash[t]{f_k^{\Psi,n}(z)}}\leq\NORM{\smash[t]{f_{k-1}^{\Psi,n}(z)}}$. 
Since this holds for all $k=1,\ldots,n$, we see
\[
\NORM{F_n(z)\Psi} = \NORM{\smash[t]{f_n^{\Psi,n}(z)} } 
\leq \ldots \leq \NORM{\smash[t]{f_1^{\Psi,n}(z)} }\leq \NORM{\Psi},
\]
as desired.
\end{proof}
The estimate proved above also yields that the $F_n$ are closable.
We will denote their closures by the same symbols in the following.
\begin{Llemm}{FnInt}
 For $z\in S$ holds the representation
\begin{equation}\labelE{FnInt}
F_n(z)A\Omega =
\frac{(z+i)^2}{2\pi i}\int_{-\infty}^{\infty}
\frac{F_n(t-i/2)A\Omega}{(t+i/2)^2(t-i/2-z)}-
\frac{F_n(t)A\Omega}{(t+i)^2(t-z)}\;\dd t.
\end{equation}
where the integrals are taken in the sense of Bochner. One further has
\begin{equation}\labelE{Fn0}
0 = \frac{1}{2\pi i}\int_{-\infty}^{\infty}
\frac{F_n(t-i/2)A\Omega}{(t+i/2)^2(t-i/2-z)}-
\frac{F_n(t)A\Omega}{(t+i)^2(t-z)}\;\dd t,  
\end{equation}
for $z\not\in\SC$.
\end{Llemm}
\begin{proof}
  By Cauchy's theorem for vector-valued functions~\cite[Theorem~3.11.3]{b:HP65},
we can write
\[
\frac{F_n(z)A\Omega}{(z+i)^2}=
\frac{1}{2\pi i}\oint\frac{F_n(\zeta)A\Omega}{(\zeta+i)^2(\zeta-z)}\;\dd\zeta,
\]
where the integral runs over a closed, positively oriented contour in $S$, 
which encloses $z$. We choose this contour to be the boundary of the rectangle
determined by the points 
$\bigl\{R-i\EPS,-R-i\EPS,-R-i(1/2-\EPS),R-i(1/2-\EPS)\bigr\}$ 
for $R>0$, $1/4>\EPS>0$. By \refL{FnBound}, the norms of the integrals over 
the paths parallel to the real line stay bounded as $R\rightarrow\infty$, 
while those of the integrals parallel to the imaginary axis vanish. Thus
\begin{gather*}
\frac{F_n(z)A\Omega}{(z+i)^2} =
\frac{1}{2\pi i}\int_{-\infty}^{\infty}
\frac{F_n(t-i(1/2-\EPS))A\Omega}{(t+i(1/2+\EPS))^2(t-i(1/2-\EPS)-z)}
\\ -\frac{F_n(t)A\Omega}{(t+i(1-\EPS))^2(t-i\EPS-z)}\;\dd t.
\end{gather*}
For $0<\EPS_0<\min\{\ABS{\IM z},\ABS{1/2-\IM z}\}$ and all 
$\EPS$ such that $0\leq\EPS\leq\EPS_0$,  the integrand is bounded in norm by
$\NORM{A}\bigm/\bigl[(1+t^2)\min\{\ABS{\IM z - \EPS_0} , \ABS{\IM z - (1/2-\EPS_0)}\}\bigr]$. 
Since moreover, in the strong sense and pointwise in $t$, 
$\lim_{\EPS\rightarrow 0}F_n(t-i\EPS)A\Omega=F_n(t)A\Omega$, and 
$\lim_{\EPS\rightarrow 0}F_n(t-i(1/2-\EPS))A\Omega=F_n(t-i/2)A\Omega$,
 the conditions for the application of the vector-valued 
Lebesgue theorem on dominated convergence~\cite[Theorem~3.7.9]{b:HP65} are given
and the desired representation follows in the limit $\EPS\rightarrow0$.
The vanishing of the second integral follows analogously.
\end{proof}
\begin{Llemm}{HoloLim}
  The strong limits $F(z)\DEF\slim_{n\rightarrow\infty}F_n(z)$, $z\in S$, 
are closable operators with common, dense domain $\OA\Omega$ (we 
denote their closures by the same symbols). The integral representation
\begin{equation}\labelE{FInt}
F(z)A\Omega =
\frac{(z+i)^2}{2\pi i}\int_{-\infty}^{\infty}
\frac{W(t-i/2)A\Omega}{(t+i/2)^2(t-i/2-z)}-
\frac{W(t)A\Omega}{(t+i)^2(t-z)}\;\dd t
\end{equation}
holds good, and the functions $F(z)A\Omega$ are holomorphic on $S$, for all 
$A\in\OA\Omega$. There exists a projection $G$ and a positive operator $\Gamma$
such that $\Gamma=G\Gamma=\Gamma G$, and ${\Gamma}^{4iz}=F(z)$ for all
$z\in S$.
\end{Llemm}
\begin{proof}
  Using \refL{FnBound}, we see that the norm of the integrand in \eqref{eq:FnInt} 
is uniformly bounded in $n$ by 
$2\NORM{A}\bigm/\bigl[(1+t^2)\min\{\ABS{\IM z},1/2-\ABS{\IM z}\}\bigr]$,
which is integrable in $t$. Furthermore, $F_n(t)A\Omega$ and $F_n(t-i/2)A\Omega$ 
converge in norm to $W(t)A\Omega$ and $W(t-i/2)A\Omega$, respectively, by 
assumptions~i) and~ii) of \refT{main}.
Thus, we can again apply Lebesgue's theorem on dominated convergence 
to infer the existence of the limits $\lim_{n\rightarrow\infty}F_n(z)A\Omega$
for all $A\in\OA$. This defines linear operators on the common, dense domain
$\OA\Omega$. Again, by the estimate of \refL{FnBound}, we have 
$F(z)A_n\Omega\rightarrow0$ if $\NORM{A_n}\rightarrow0$, and therefore the
$F(z)$ are closable. The validity of \refE{FInt} is then clear.
Since the bound noted above is uniform in $n$, and all the functions
$F_n(z)A\Omega$ are holomorphic in $S$, we can apply the Stieltjes--Vitali
theorem~\cite[Theorem~3.14.1]{b:HP65} to deduce the stated holomorphy of $F(z)A\Omega$.
We now consider the operators $F(-is)$, $0<s<1/2$. Using the same properties
of $\Delta$, $E$, one sees that these operators are self-adjoint, and in fact,
positive: Namely, the limits are densely defined, symmetric and closable
operators, and an analytic vector for $\Delta^{1/2}$ is also analytic
for $F(-is)$, $0<s<1/2$. Thus the $F(-is)$ possess a common, dense
set of analytic vectors. Under these circumstances,  the $F(-is)$ are 
essentially self-adjoint, and we denote their unique, self-adjoint extension
by the same symbol. 
We now follow~\cite{MS77} to show that the functional equation 
$F(-i(s+t))=F(-is)F(-it)$ holds for $s$, $t>0$ such that 
$s+t<1/2$. To this end, consider first the case that $s$ and $t$ are
rationally related\ie there exist $p$, $q\in\NN$ such that
\[
\frac{s+t}{r(p+q)}=\frac{s}{rp}=\frac{t}{rq}, \quad\text{for all }r\in\NN.
\]
Then
\[
\left[E\Delta^{\frac{s+t}{r(p+q)}}E\right]^{r(p+q)}A\Omega=
\left[E\Delta^{\frac{s}{rp}}E\right]^{rp}
\left[E\Delta^{\frac{t}{rq}}E\right]^{rq}A\Omega,\quad A\in\OA,
\]
from which the claim follows in the limit $r\rightarrow\infty$.
The general case follows since $F(-is)A\Omega$ is holomorphic
and therefore also strongly continuous in $s$ for all $A\in\OA$.
Now set ${\Gamma}=F(-i/4)$. By the spectral 
calculus for unbounded operators~\cite[Section~5.6]{b:KR83/86}, 
the positive powers
$\Gamma^\sigma$ exist for $0<\sigma\leq1$, and are positive operators
with domain containing the common, dense domain $\OA\Omega$. They
satisfy the functional equation 
${\Gamma}^{\sigma+\tau}=
{\Gamma}^{\sigma}\CDOT{\Gamma}^{\tau}$
for $\sigma$, $\tau>0$ such that $\sigma+\tau\leq1$, and where
$\CDOT$ denotes the closure of the operator product. The solution
to this functional equation with initial condition 
${\Gamma}=F(-i/4)$ is unique and thus it follows
${\Gamma}^\sigma=F(-i\sigma/4)$, since the operators $F$ 
satisfy the same functional equation, and all operators in
question depend continuously on $\sigma$, in the strong sense
when applied to the common core $\OA\Omega$. For $1/4\leq s< 1/2$ 
we have $F(-is)=F(-i/4)F(-i(s-1/4))={\Gamma}F(-i(s-1/4))=
{\Gamma}{\Gamma}^{4s-1}={\Gamma}^{4s}$,
which finally shows the identity $F(-is)=\Gamma^{4s}$ for
$0<s<1/2$. Now, for every $A\in\OA$, 
${\Gamma}^{4iz}A\Omega$ extends to a holomorphic
function on $S$ which coincides with $F(z)A\Omega$ on the segment 
$\{-is\mid 0<s<1/2\}$ as we have just seen. 
The identity theorem for vector-valued, 
holomorphic functions~\cite[Theorem~3.11.5]{b:HP65} then implies
${\Gamma}^{4iz}A\Omega=F(z)A\Omega$, $z\in S$ and all 
$A\in\OA$. Thus ${\Gamma}^{4iz}=F(z)$ holds on $S$ as an 
identity of densely defined, closed operators. Setting $G=P([0,\infty))$,
where $P$ is the spectral resolution of the identity for ${\Gamma}$,
we see that we can write ${\Gamma}=G\Gamma =\Gamma G$, concluding this
proof.
\end{proof}
\begin{Llemm}{Boundary}
  It holds $G=E$, and $W(t)=E\Gamma^{4it}E$,
for all $t$.
\end{Llemm}
\begin{proof}
Using~\eqref{eq:FInt} we can write, adding a zero contribution to that 
integral representation,
\begin{multline*}
  \SPROD{B\Omega}{F(t-i\eta)A\Omega} =\\
\frac{(t+i-i\eta)^2}{2\pi i}\int_{-\infty}^{\infty}\;\dd s\Biggl\{
\frac{\SPROD{B\Omega}{W(s-i/2)A\Omega}}{(s+i/2)^2(s-t-i/2+i\eta)}-
\frac{\SPROD{B\Omega}{W(s)A\Omega}}{(s+i)^2(s-t+i\eta)}-\\
-\frac{\SPROD{B\Omega}{W(s-i/2)A\Omega}}{(s+i/2)^2(s-t-i/2-i\eta)}+
\frac{\SPROD{B\Omega}{W(s)A\Omega}}{(s+i)^2(s-t-i\eta)}
\Biggr\},
\end{multline*}
where the integral over the last two terms is zero, as can be seen
from~\eqref{eq:Fn0} and the same arguments that were used to 
derive~\eqref{eq:FInt}. This yields
\[
= 
\frac{(t+i-i\eta)^2}{\pi}\int_{-\infty}^{\infty}
\frac{\eta\cdot\SPROD{B\Omega}{W(s-i/2)A\Omega}}{(s+i/2)^2((s-t-i/2)^2+\eta^2)}-
\frac{\eta\cdot\SPROD{B\Omega}{W(s)A\Omega}}{(s+i)^2((s-t)^2+\eta^2)}
\;\dd s.
\]
As $\eta\rightarrow0_+$, the first term under the integral vanishes,
while the second reproduces the integrable function
$\SPROD{A\Omega}{W(t)B\Omega}/(t+i)^2$ as the boundary
value of its Poisson transformation. Thus we have seen
\[
\lim_{\eta\rightarrow0_+}
\SPROD{A\Omega}{F(t-i\eta)B\Omega}=\SPROD{A\Omega}{W(t)B\Omega},
\]
for given $A$, $B\in\OA$, and almost all $t\in\RR$.
Since the integral is uniformly bounded in $\eta$,
the boundary value of this Poisson transformation
is continuous in $t$, see\eg~\cite[Section~5.4]{b:BRE65}.
The same holds for \SPROD{A\Omega}{W(t)B\Omega} by assumption~i)
of \refT{main}, and therefore the limiting identity at $\eta=0$
follows for all $t$. 
On the other hand, since $G\Gamma^{4it}G$ is strongly continuous in $t$,
we have $\lim_{\eta\rightarrow0_+}
\SPROD{A\Omega}{F(t-i\eta)B\Omega}=
\SPROD{A\Omega}{G\Gamma^{4it}GB\Omega}$
for all $t$.
Thus, the identity of bounded operators
$W(t)=G\Gamma^{4it}G$ holds for all $t$. By assumption we have
$\wlim_{t\rightarrow0}W(t)=E$, thus 
$W(s)W^\ast(s)=G\Gamma^{4is}\Gamma^{-4is}G=G$ implies $G=E$.
\end{proof}
\begin{Llemm}{auto}
  The action 
$\tau_t^E\colon\OA_E\ni A_E\longmapsto\tau^E_t(A_E)=\Gamma^{4it}A_E\Gamma^{-4it}$
is a strongly continuous group of automorphisms of $\OA_E$.
\end{Llemm}
\begin{proof}
  For $A_E=EAE\in\OA_E$ we have $E\Delta^{it/n}E A_E E\Delta^{-it/n}E=
E\sigma_{t/n}(A_E)E$, 
where $\sigma$ is the modular group of $(\OA,\Omega)$, and this 
shows $F_n(t)A_EF_n(-t)\in\OA_E$ for all $n$. 
Since $\OA_E$ is weakly closed
and $F_n(t)A_EF_n(-t)$ converges strongly, and therefore also weakly, by 
assumption~i) of \refT{main}, it converges to an element of $\OA_E$.
Since $\NORM{F_n(t)A_EF_n(-t)}\leq\NORM{A_E}$ for all $n$,
the limit mapping is continuous
on $\OA_E$. By \refL{Boundary}, it equals $\tau^E_t$, as defined above,
for all $t$. Since $\Gamma^{4it}$ is a strongly continuous group of 
unitary operators on $E\HS$, the assertion follows.
\end{proof}
\begin{proof}[Proof of \refT{main} and \refC{modular-identify}]
We note first, that $W(-t)=W(t)^\ast$ can be seen by direct methods
as in~\cite{MS77}. Secondly, since $\tau^E$ is an automorphism group
of $\OA_E$, it follows by definition of $\HS_E$, that the $W(t)$ leave
that subspace invariant and thus form a unitary group on it.
The stated analyticity properties of $W$ are contained in the conlusions
of Lemmata~\ref{lemm:HoloLim} and~\ref{lemm:Boundary}.

Let us now turn to the identification of $W$ with 
the modular group of the pair $(\OA_E,\Omega_E)$.
An argument as was used in the proof of \refL{Boundary}
shows
\[
\SPROD{A\Omega}{\smash{\Gamma^{4i(t-i/2)}}B\Omega}=
\lim_{\eta\rightarrow1/2_-}\SPROD{A\Omega}{F(t-i\eta)B\Omega}=
\SPROD{A\Omega}{W(t-i/2)B\Omega},
\]
for given  $A$, $B\in\OA$, and almost all $t\in\RR$.
Additionally, as mentioned in the proof of
\refL{Boundary}, the boundary value of the Poisson 
transformation $\Gamma^{4i(t-i/2)}$ is weakly continuous on $\OA\Omega$.
From this, the density of $\OA\Omega$ in \HS,  
and assumption iii) of \refT{main} it follows for $A_E$, $B_E\in\OA_E$:
\begin{align*}
\SPROD{\smash{\Gamma^2 A_E\Omega_E}}{\smash{\Gamma^2 A_E\Omega_E}}&=
\lim_{t\rightarrow0}
\SPROD{\smash{\Gamma^{4i(-t-i/2)}B_E\Omega_E}}{\smash{\Gamma^{4i(t-i/2)}A_E\Omega_E}}\\
&=\lim_{t\rightarrow0}\SPROD{\smash{W(-t-i/2)A_E\Omega_E}}{\smash{W(t-i/2)A_E\Omega_E}}\\ &
=\SPROD{\smash{\Delta^{1/2}A_E\Omega_E}}{\smash{\Delta^{1/2}A_E\Omega_E}},\\
\intertext{where the weak continuity has been used in the first and
the identity above in the second step.
Now by the modular condition satisfied by
$\Delta$, this becomes}
&=\SPROD{B_E^\ast\Omega_E}{A_E^\ast\Omega_E}.
\end{align*}
This is the modular condition for the automorphism group $\tau^E$
with respect to $(\OA_E,\Omega_E)$.  The assertion of the theorem 
follows by uniqueness of the modular 
group~\cite[Theorem~9.2.16]{b:KR83/86} 
and the preceding three Lemmata.
\end{proof}
\section{Conclusions}
Let us comment a bit on the status of \refT{main}. 
It is stronger than the result of~\cite{MS77} in that it
generalizes it to KMS-states of $W^\ast$-dynamical systems at 
inverse temperatures $0<\beta\leq\infty$. This is exactly
the framework in which a strip of analyticity of width
$\beta$ above (or below, depending on convention) the
real axis exists, which is the sole condition needed to
apply the methods of complex analysis used extensively to
prove \refT{main}. This shows that Misra's and Sudarshan's
theorem~\cite{MS77} can be extended to cases in which
the Hamiltonian is not lower semibounded, but in which 
its negative spectral parts are ``exponentially damped''
(see~\cite[Section~V.2.1]{b:HAA92} for the precise meaning 
of these notions). This is in contrast to the counterexample
in~\cite{MS77}, where those authors state that lower semiboundedness
is essential. That counterexample involves the unitary group 
generated by the momentom operator, which does not fulfill
any requirement of exponential damping of the negative spectral
part, and thus violates our analyticity assumptions.
See also the discussion in~\cite[Section~3]{HW86}, where
it is noted that lower semiboundedness does not seem to be important for
the Zeno effect in general

But more interestingly,
\refC{modular-identify} identifies the induced Zeno dynamics uniquely, 
as already mentioned in the introduction. 
For this, however, we needed the additional assumption~iii) in \refC{modular-identify},
and this assumption is not a simple consequence of the modular 
condition of the original system. But notice that we did not use
assumption~iii) until after \refL{auto}, so that in any case we get
an automorphic Zeno dynamics on $\OA_E$, which however might differ
from the modular dynamics, depending on the choice of state on $\OA_E$,
but has the analyticity properties needed to check the modular
condition for a given state as in \refC{modular-identify}. 
Assumption~iii) is therefore only a formal condition for $\Omega_E$
to be a $\tau^E$-KMS state, essentially a straightforward transcription
of the modular condition itself.
One simple example of an equilibrium state for
the Zeno dynamics can be given in the case when the projector $E$ commutes
with the initial dynamics\ie $[E,\Delta]=0$. Then the Zeno dynamics
simplifes to $W(t)=E\Delta^{it}E$ and the state 
$\Omega_E\DEF E\Omega/\NORM{E\Omega}$ satisfies condition~iii).
Other, less trivial, examples can arise from Gibbs equilibria, as is shown
in~\cite{AUS02B}.

As already remarked in~\cite{MS77}, 
the relatively strong assumptions of the theorem might be difficult
to prove in concrete cases. It seems more likely that in studying physical
models one would rather identify the nature of the induced Zeno dynamics
directly, as for example in~\cite{FGMPS00,FPSS01}. Therefore \refT{main} 
and \refC{modular-identify} are
to be considered as a mathematical Gedankenexperiment, which might 
be helpful in guiding physical intuition.

However, some relaxations of the assumptions of \refT{main} are 
possible in special cases~\cite{MS77}:
Firstly, if the theory contains a CPT-operator, the conclusions of
\refT{main} already follow if one assumes the convergence of the
limits defining $W(t)$, $W(t-i/2)$ only for positive times.
Secondly, if we restrict attention to von Neumann algebras \OA that allow a
faithful representation on a \emph{separable} Hilbert space \HS, we 
can drop the assumption of weak continuity of the limits $W(t)$, 
$W(t-i/2)$ used in the proof of \refL{Boundary}, 
by the argument given in~\cite{MS77}: First show that
$\lim_{\eta\rightarrow0_+}
\SPROD{A\Omega}{F(t-i\eta)B\Omega}=\SPROD{A\Omega}{W(t)B\Omega}$
for $t$ outside an exceptional null set $\mathcal{N}_{A,B}\subset\RR$. 
Then, there exists a countable set $\mathcal{C}\subset\OA$ such that 
$\mathcal{C}\Omega$ is dense in \HS and the countable union
$\bigcup_{A,B\in\mathcal{C}}\mathcal{N}_{A,B}$ is still a null set.
One easily shows that the limit relation holds true outside this set
in the weak sense on \HS and proceeds from there using the strong 
continuity of $G\Gamma^{4it}G$ as in~\cite{MS77}. 

An application, 
to simple examples of quantum statistical mechanics, like
spin systems, seems possible. There, if the projection $E$
projects to some pure state
outside a bounded region, one is in a
case where one would presume the Zeno
dynamics to exist. One could expect to find the dynamics of the
bounded region of the spin system (a matrix model) with appropriate
boundary conditions, to be determined by the action of $E$ on the
boundary layer (see~\cite{FW95} for a complementary discussion).
This, and further applications in the context of quantum statistical
mechanics are found in a subsequent paper~\cite{AUS02B}.

There will also appear a subtle
point in physical applications: When considering models
one usually deals with $C^\ast$-dynamical systems rather
than $W^\ast$-dynamical ones\ie the relevant algebras are
norm rather than weakly closed. For these, the set of KMS-states
at given inverse temperature is in general a nonempty, weak $\ast$-closed,
convex subset of the state space~\cite[Section~5.3.2]{b:BR79/81}.
Thus, one always has to choose a KMS-state and an associated
representation to work within. If one fixes a KMS-state, 
say $\omega$ with vector representative $\Omega$,  
and wants to consider the induced Zeno dynamics determined by a projector 
$E$, the restricted subsytem $\OA_E$ will still have in general a 
multitude of KMS-states of its own. The point is that it is not
{\it a priori} clear that the Zeno dynamics will leave a chosen
state $\omega_E$ invariant. It may happen that the Zeno dynamics transforms the
KMS-states of $\OA_E$ into each other in a nontrivial way.
In this case the Zeno dynamics is not unitary\ie reversible.
This is a reflection of the same problem appearing in the
quantum mechanical context~\cite[Section~5]{FGMPS00}.
This cannot happen, however, within the context of \refT{main},
for if the Zeno dynamics satisfies condition iii), then invariance
of $\Omega_E$ follows directly~\cite[Proposition ~5.3.3]{b:BR79/81}.
But even if the equilibrium condition~iii) of that theorem
is not satisfied, \refL{auto} still assures that the induced dynamics is 
unitary. For these problems, see also the recent approach of 
Gustafson~\cite{GUS02}, who tackles the associated problem of 
selfadjointness of the generator of the Zeno dynamics
(In~\cite{GUS02}, one also finds some important
remarks on the history of the subject).

Finally, we remark that we were in part motivated
by the proposal to take the modular flow of observable algebras 
as a definition of physical time~\cite{ROV93A,ROV93B,CR94}, 
for example on a generally relativistic background, when the usual concepts 
are less useful. However, the relation between modular groups
and space-time is an intricate one, as has local 
quantum physics taught us~\cite{BY99,BOR00}, and this
thread of thought might be just at its beginning.
%

%

\begin{thebibliography}{10}
\providecommand{\noopsort}[1]{} \providecommand{\singleletter}[1]{#1}
\providecommand{\bysame}{\leavevmode\hbox to3em{\hrulefill}\thinspace}
\bibitem{AA90}
J.~Anandan and Y.~Aharanov, \emph{{G}eometry of {Q}uantum {E}volution}, Phys.
  Rev. Lett. \textbf{65} (1990), no.~14, 1697--1700.

\bibitem{BOR00}
H.~J. Borchers, \emph{{O}n revolutionizing quantum field theory with {T}omita's
  modular theory}, J. Math. Phys. \textbf{41} (2000), no.~6, 3604--3673.

\bibitem{BY99}
H.~J. Borchers and J.~Yngvason, \emph{{M}odular groups of quantum fields in
  thermal states}, J. Math. Phys. \textbf{40} (1999), no.~2, 601--624,
  \href{http://xxx.lanl.gov/abs/math-ph/9805013}{\texttt{math-ph/9805013}}.
  MR \href{http://www.ams.org/mathscinet-getitem?mr=2000d:81085}{2000d:81085}

\bibitem{b:BR79/81}
Ola Bratteli and Derek~W. Robinson, \emph{{O}perator {A}lgebras and {Q}uantum
  {S}tatistical {M}echanics}, vol. I~\&~II, Springer-{V}erlag, Berlin,
  {H}eidelberg, {N}ew {Y}ork, 1979~\&~1981.

\bibitem{b:BRE65}
Hans Bremermann, \emph{Distributions, {C}omplex {V}ariables, and {F}ourier
  {T}ransforms}, Addison--Wesley Publishing Company, Reading, Massachusetts,
  1965.

\bibitem{CR94}
Alain Connes and Carlo Rovelli, \emph{{V}on {N}eumann algebra automorphisms and
  time-ther\-mo\-dy\-na\-mics relation in generally covariant quantum theories},
  Classical Quantum Gravity \textbf{11} (1994), no.~12, 2899--2917.

\bibitem{FGMPS00}
P.~Facchi, V.~Gorini, G.~Marmo, S.~Pascazio, and E.~C.~G. Sudarshan,
  \emph{{Q}uantum {Z}eno dynamics}, Phys. Lett. A \textbf{275} (2000), 12--19,
  \href{http://xxx.lanl.gov/abs/quant-ph/0004040}{\texttt{quant-ph/0004040}}.

\bibitem{FNP01}
P.~Facchi, H.~Nakazato, and S.~Pascazio, \emph{{F}rom the {Q}uantum {Z}eno to
  the {I}nverse {Q}uantum {Z}eno {E}ffect}, Phys. Rev. Lett. \textbf{86}
  (2001), no.~13, 2699--2703.

\bibitem{FPSS01}
P.~Facchi, S.~Pascazio, A.~Scardicchio, and L.~S. Schulman, \emph{{Z}eno
  dynamics yields ordinary constraints}, Phys. Rev. A \textbf{65} (2001),
  012108,
  \href{http://xxx.lanl.gov/abs/quant-ph/0101037}{\texttt{quant-ph/0101037}}.

\bibitem{FW95}
M. Fannes, and R. F. Werner, \emph{Boundary Conditions for Quantum Lattice Systems},
Helv. Phys. Acta \textbf{68} no.~7-8 (1995), 635--657.

\bibitem{GUS02}
Karl Gustafson, \emph{A Zeno Story}, preprint, March 2002.
\href{http://xxx.lanl.gov/abs/quant-ph/0203032}{\texttt{quant-ph/0203032}}.

\bibitem{b:HAA92}
Rudolf Haag, \emph{Local Quantum Physics}, 
Springer-{V}erlag, Berlin, {H}eidelberg, {N}ew {Y}ork, 1992.

\bibitem{b:HP65}
Einar Hille and Ralph~S. Phillips, \emph{Functional {A}nalysis and
  {S}emigroups}, revised ed., AMS Colloquium Publications, vol. XXXI, American
  {M}athematical {S}ociety, American {M}athematical {S}ociety, 1965.

\bibitem{HW86}
D.~Home and M.~A.~B. Whitaker, \emph{Reflections on the quantum Zeno paradox},
J. Phys. A \textbf{19} (July 1986), 1847--1854.

\bibitem{HW98}
\bysame, \emph{{Q}uantum {Z}eno {E}ffect: {R}elevance for
  local realism, macroscopic realism, and non-invasive measurability at the
  macroscopic level}, Phys. Lett. A \textbf{239} (1998), 6--12.

\bibitem{HNNPR98}
Zdenek Hradil, Hiromichi Nakazato, Mikio Namiki, Saverio Pascazio, and Helmut
  Rauch, \emph{{I}nfinitely frequent measurements and quantum {Z}eno effect},
  Phys. Lett. A \textbf{239} (1998), 333--338.

\bibitem{b:KR83/86}
Richard~V. Kadison and John~R. Ringrose, \emph{Fundamentals of the {T}heory of
  {O}perator {A}lgebras}, vol. I~\&~II, Academic {P}ress, New York, London,
  Paris, San Diego, 1983~\&~1986.

\bibitem{MS77}
B.~Misra and E.~C.~G. Sudarshan, \emph{{T}he {Z}eno's paradox in quantum
  theory}, J. Math. Phys. \textbf{18} (1977), no.~4, 756--763.

\bibitem{NNP96}
Hiromichi Nakazato, Mikio Namiki, and Saverio Pascazio, \emph{{T}emporal
  behavior of quantum mechanical systems}, Internat. J. Modern Phys. B
  \textbf{10} (1996), 247,
  \href{http://xxx.lanl.gov/abs/quant-ph/9509016}{\texttt{quant-ph/9509016}}.

\bibitem{PL98}
A.~K. Pati and S.~V. Lawande, \emph{{G}eometry of the {H}ilbert space and the
  quantum {Z}eno effect}, Phys. Rev. A \textbf{58} (1998), no.~2, 831--835.

\bibitem{ROV93A}
Carlo Rovelli, \emph{{S}tatistical mechanics of gravity and the thermodynamical
  origin of time}, Classical Quantum Gravity \textbf{10} (1993), no.~8,
  1549--1566.

\bibitem{ROV93B}
\bysame, \emph{{T}he statistical state of the universe}, Classical Quantum
  Gravity \textbf{10} (1993), no.~8, 1567--1578.

\bibitem{AUS02B}
  Andreas U. Schmidt, 
  \ifpdf%
   \href{http://www.math.uni-frankfurt.de/~aschmidt/paper/zeno_qstat_amsart.pdf}{%
   \emph{Zeno Dynamics in Quantum Statistical Mechanics}},
  \else
   \emph{Zeno Dynamics in Quantum Statistical Mechanics},
  \fi
  Preprint, Dipartimento di Fisica, Universit\'a di Pisa, Italy, July 2002,
  \href{http://xxx.lanl.gov/abs/math-ph/0207013}{\texttt{math-ph/0207013}}.

\bibitem{b:WIC95}
David Wick, \emph{The {I}nfamous {B}oundary. {S}even {D}ecades of {H}eresy in
  {Q}uantum {P}hysics.}, Birkh{\"a}user--{V}erlag, Boston, Basel, Stuttgart,
  1995.

\bibitem{WBT01}
Chr. Wunderlich, Chr. Balzer, and P.~E. Toschek, \emph{{E}volution of an {A}tom
  {I}mpeded by {M}easurement: {T}he {Q}uantum {Z}eno {E}ffect}, Z. Naturforsch.
  \textbf{56a} (2001), 160--164.
  \href{http://xxx.lanl.gov/abs/quant-ph/0108040}{\texttt{quant-ph/0108040}}.

\end{thebibliography}
\end{document}